\documentclass[%
reprint,
superscriptaddress,
showpacs,
amsmath,amssymb,
prc,
floatfix, ]%
{revtex4-1}

\usepackage{color}

\usepackage{graphicx}
\usepackage{dcolumn}
\usepackage{bm}
\usepackage[dvipdfmx,bookmarks=true,colorlinks,%
            citecolor=blue,linkcolor=blue,anchorcolor=blue,filecolor=blue,urlcolor=blue,%
           ]{hyperref}

\allowdisplaybreaks

\begin{document}


\title{Energy-Dependence of Nucleus-Nucleus Potential and Friction Parameter in Fusion Reactions}

\author{Kai Wen}
 \affiliation{State Key Laboratory of Theoretical Physics,
              Institute of Theoretical Physics, Chinese Academy of Sciences,
              Beijing 100190, China}
\author{Fumihiko Sakata}
 \affiliation{Institute of Applied Beam Science, Graduate School of Science and Technology,
              Ibaraki University, Mito 310-8512, Japan}
 \affiliation{State Key Laboratory of Theoretical Physics,
              Institute of Theoretical Physics, Chinese Academy of Sciences,
              Beijing 100190, China}
\author{Zhu-Xia Li}
 \affiliation{China Institute of Atomic Energy, Beijing 102413, China}
\author{Xi-Zhen Wu}
 \affiliation{China Institute of Atomic Energy, Beijing 102413, China}
\author{Ying-Xun Zhang}
 \affiliation{China Institute of Atomic Energy, Beijing 102413, China}
\author{Shan-Gui Zhou}
 \email{sgzhou@itp.ac.cn}
 \affiliation{State Key Laboratory of Theoretical Physics,
              Institute of Theoretical Physics, Chinese Academy of Sciences,
              Beijing 100190, China}
 \affiliation{Center of Theoretical Nuclear Physics, National Laboratory
              of Heavy Ion Accelerator, Lanzhou 730000, China}
 \affiliation{Center for Nuclear Matter Science, Central China Normal University,
              Wuhan 430079, China}

\date{\today}

\begin{abstract}
Applying a macroscopic reduction procedure on the improved quantum molecular
dynamics (ImQMD) model, the energy dependences of the nucleus-nucleus potential,
the friction parameter, and the random force characterizing a one-dimensional
Langevin-type description of the heavy-ion fusion process are investigated.
Systematic calculations with the ImQMD model show that the fluctuation-dissipation
relation found in the symmetric head-on fusion reactions at energies just above
the Coulomb barrier fades out when the incident energy increases.
It turns out that this dynamical change with increasing incident energy
is caused by a specific behavior of the friction
parameter which directly depends on the microscopic dynamical process, i.e.,
on how the collective energy of the relative motion is transferred into
the intrinsic excitation energy.
It is shown microscopically that the energy dissipation in the fusion process
is governed by two mechanisms:
One is caused by the nucleon exchanges between two fusing nuclei,
and the other is due to a rearrangement of nucleons in the intrinsic system.
The former mechanism monotonically increases the dissipative energy and
shows a weak dependence on the incident energy,
while the latter depends on both the relative distance between two fusing nuclei
and the incident energy.
It is shown that the latter mechanism is responsible for the energy dependence of
the fusion potential and explains the fading out of the fluctuation-dissipation
relation.

\end{abstract}


\maketitle


\section{Introduction}

Energy dissipation process observed in heavy-ion reactions ranging from
deep-inelastic collisions to sub-barrier fusions has been described by various
macroscopic transport models~\cite{%
Bjornholm1982_NPA391-471,
Frobrich1998_PR292-131,
Adamian1998_NPA633-409,
Shen2002_PRC66-061602R,
Zagrebaev2007_JPG34-2265,
Zubov2009_PPN40-847,
Li2010_NPA834-353c,
Aritomo2012_PRC85-044614,
Siwek-Wilczynska2012_PRC86-014611,
Nasirov2011_PRC84-044612,
Wang2012_PRC85-041601R,
Liu2013_PRC87-034616,
Nasirov2013_EPJA49-147,
Gontchar2014_PRC89-034601}. 
Although many of these macroscopic models are very successful in predicting the cross
section of compound nucleus formation, their microscopic foundation, e.g.,
how the colliding two nuclei fuse and how the relative kinetic energy dissipates
into the intrinsic energy, still requires further research.

Besides these macroscopic transport models, many microscopic approaches
like the time-dependent Hartree-Fock (TDHF) theory~\cite{Bonche1976_PRC13-1226,
Koonin1980_PPNP4-283,
Brink1981_PRC24-144,
Negele1982_RMP54-913,
Umar1986_PRL56-2793,
Guo2007_PRC76-014601, Guo2008_PRC77-041301R,
Washiyama2008_PRC78-024610,
Washiyama2009_PRC79-024609,
Ayik2009_PRC79-054606,
Simenel2012_EPJA48-152,
Simenel2013_PRC88-064604,
Dai2014_PRC, Dai2014_SciChinaPMA57-1618},
the many-body correlation transport theory~\cite{Wang1985_AoP159-328},
and
the quantum molecular dynamics (QMD)~\cite{Aichelin1991_PR202-233, Sorge1989_NPA498-567},
the antisymmetrized molecular dynamics~\cite{Ono1999_PRC59-853, Horiuchi1994_NATO-ASI335-215},
and
the fermion molecular dynamics~\cite{Feldmeier2000_RMP72-655}
models
have been developed.
It is worthwhile mentioning that, among the TDHF approaches,
there have been proposed new methods in recent years like the dissipative-dynamics (DD)
TDHF \cite{Washiyama2008_PRC78-024610, Washiyama2009_PRC79-024609}
and the density-constrained (DC) TDHF \cite{Umar2006_PRC74-021601R,
Umar2007_PRC76-014614} theories, which intend to explore how the macroscopic
collective behavior of two colliding nuclei described by the Langevin-type equation
appears as a result of huge dimensional microscopic dynamics.
With the aid of these methods, the incident energy dependences of the macroscopic potential,
the collective inertia parameter as well as the friction force have been derived
from microscopic TDHF simulations and the discussions of which are made
in relation with the internal structure change of the colliding nuclei.

As is widely accepted, the above subject of heavy-ion fusion
process is related to a more general problem, i.e., how to get a better
understanding on various macroscopic collective behavior of non-equilibrium
systems in a deeper level, which is currently studied in many fields of
sciences, e.g., in physical and chemical as well as biological systems,
by exploiting state-of-the-art {\it ab-initio} numerical simulations of
the molecular dynamics model 
\cite{Karplus2002_NSMB9-646,*Karplus2002_NSMB9-788, Carloni2002_ACR35-455}.

Recently, we have proposed a macroscopic reduction procedure based on
the improved quantum molecular dynamics (ImQMD) model,
aiming at exploring how the macroscopic Langevin-type equation emerges
out of the microscopic dynamics in such a finite quantum many-body
system as an atomic nucleus \cite{Wen2013_PRL111-012501}.
We found that the dissipation dynamics of the relative motion between
two fusing nuclei is caused by a non-Gaussian distribution of the random force.
A clear non-Markovian effect was also observed in the time correlation function
of the random force.
Note that the non-Gaussian fluctuation is a general feature in non-equilibrium
systems ranging from the cosmoscopic \cite{Maldacena2003_JHEP05-013,
Matarrese1986_ApJ310-L21},
mesoscopic \cite{Akimoto2011_PRL107-178103, Laakso2012_PRL108-067002},
to microscopic systems \cite{Stephanov2009_PRL102-032301}.
Furthermore, as discussed in Ref.~\cite{Pachon2014_arXiv1401.1418},
the non-Markovian dynamics were relevant in many fields and applications.

In this paper, we further develop our macroscopic reduction procedure
applicable to the ImQMD model.
An extension is based on a new method~\cite{Sakata2014_in-prep} which projects
the effects of intrinsic degrees of freedom onto the collective subspace
by transforming the time-scale into a macroscopic collective space-scale.
With the aid of this macroscopic scale transformation, we explicitly extract
the effects of intrinsic degrees of freedom in a convenient form
that is appropriate for discussing the dynamical change of the fusion reaction
from the adiabatic regime to the diabatic one.
Exploiting the representation obtained after the transformation,
one may clearly discuss how the dynamical role
of the intrinsic system changes when the incident energy increases and how the
energy transfer from collective motion to the intrinsic one,
which is found to be
carried out through both nucleon transfer between two fusing nuclei and
the rearrangement effects in the intrinsic system.

The paper is organized as follows.
In Sec.~\ref{sec:ImQMD}, we briefly introduce the ImQMD model
and recapitulate the macroscopic reduction procedure of the ImQMD model
proposed in Ref.~\cite{Wen2013_PRL111-012501}.
Applying the macroscopic reduction method of the ImQMD model to $^{90}$Zr+$^{90}$Zr
head-on fusion reaction, in Sec.~\ref{sec:energy_dep},
numerical results on the incident energy dependences of the random force,
the fluctuation-dissipation relation, and the fusion potential are discussed.
In Sec.~\ref{sec:micro_dyn}, we introduce various macroscopic quantities
by further developing the macroscopic reduction procedure suitable for
exploring the macroscopic dissipative motion.
Applying these new quantities on the numerical simulations of the ImQMD model,
it is clearly shown that the energy dissipation is characterized by
two competitive microscopic processes.
The energy dependence of the fluctuation-dissipation relation and
that of the fusion potential are consistently explained as a result of
these competitive factors.
Concluding remarks are given in Sec.~\ref{sec:summary}.
In the appendix, we give the derivation of the energy dissipation due to
the nucleon exchange.

\section{\label{sec:ImQMD}ImQMD model and the Macroscopic Reduction Procedure}

\subsection{ImQMD}

Like in the original QMD model~\cite{Aichelin1991_PR202-233} and in various
modern versions of QMD models~\cite{Wei2004_PLB586-225, Li2009_SciChinaG52-1530,
Guo2012_SciChinaPMA55-252, Wang2012_SciChinaPMA55-2407, Feng2008_NPA802-91,
Kumar2012_PRC86-051601R, Zhu2013_NPA915-90, Feng2013_NPA919-32}, a trial wave function
for a nucleus in the ImQMD model~\cite{Wang2002_PRC65-064608, Wang2004_PRC69-034608,
Wang2006_PRC74-044604, Zhao2009_PRC80-054607, Jiang2013_PRC88-044611}
is restricted within a parameter space $\{\textbf{r}_{j}, \textbf{p}_{j}\}$,
where $\textbf{r}_{j}$ and $\textbf{p}_{j}$ are mean values of position and
momentum operators of the $j$th nucleon expressed by a Gaussian wave packet,
and the total wave function is a direct product of these wave functions of Gaussian form.
With the aid of a Wigner transformation, a nucleus composed of distinguishable
$N$ nucleons is characterized by the following one-body phase space distribution
function,
\begin{eqnarray}
f(\textbf{r}, \textbf{p}) = \sum_{i}\frac{1}{(\pi \hbar)^{3}}
                            \exp\left[-\frac{(\textbf{r}-\textbf{r}_{i})^{2}}{2\sigma^{2}_{r}}
                            - \frac{2\sigma^{2}_{r}}{\hbar^{2}}(\textbf{p}-\textbf{p}_{i})^{2}\right],\label{eq1-1}
\end{eqnarray}
and a density distribution function is expressed as
\begin{eqnarray}
\rho(\textbf{r}) &=& \int f(\textbf{r}, \textbf{p})d\textbf{p} \nonumber\\
                 &=&\sum_{i} \frac{1}{(2\pi \sigma^{2}_{r})^{3/2}}
                 \exp\left[-\frac{(\textbf{r}-\textbf{r}_{i})^{2}}{2\sigma^{2}_{r}}\right], \label{eq1-2}
\end{eqnarray}
where $\sigma_{r}$ represents the spatial width of the nucleon wave packet.

The time evolution of the system in question is governed by a set of canonical equations of motion
\begin{eqnarray}
 \dot{\textbf{p}}_{i} =-\frac{\partial H}{\partial \textbf{r}_{i}},\quad  \dot{\textbf{r}}_{i}
            =\frac{\partial H}{\partial \textbf{p}_{i}}, \label{eq1-3}
\end{eqnarray}
which is derived from the time-dependent variational principle~\cite{Aichelin1991_PR202-233,
Ono1999_PRC59-853,Feldmeier2000_RMP72-655}.
Upon that, collisions are included in the ImQMD simulations and
the scattering angle of a single nucleon-nucleon collision is randomly chosen in such a way
that the distribution of the scattering angles of all collisions agrees with the measured
angular distribution for elastic and inelastic collisions~\cite{Aichelin1991_PR202-233}.
The Hamiltonian $H$ consists of the kinetic energy and an effective
interaction potential energy,
\begin{eqnarray}
H=T+U_{\rm loc}+U_{\rm Coul}.\label{eq1-4}
\end{eqnarray}
Here, $U_{\rm Coul}$ denotes the Coulomb energy and the nuclear potential
energy $U_{\rm loc}$ is expressed as
\begin{eqnarray}
U_{\rm loc}&=&\int V_{\rm loc}[\rho(\textbf{r})]d\textbf{r},\cr
V_{\rm loc}[\rho]&=&\frac{\alpha}{2}\frac{\rho^{2}}{\rho_{0}}
              +\frac{\beta}{\gamma+1}\frac{\rho^{\gamma+1}}{\rho^{\gamma}_{0}}
              +\frac{g_{\rm sur}}{2\rho_{0}}(\nabla\rho)^{2}\nonumber\\
              &+&\frac{C_{s}}{2\rho_{0}}\left[(\rho^{2}- \kappa_{s}(\nabla\rho)^{2}  \right]\delta^{2}
              +g_{\tau}\frac{\rho^{\eta + 1}}{\rho_{0}^{\eta}}.\label{eq1-5}
\end{eqnarray}
In the present paper, $V_{\rm loc}[\rho]$ is defined by applying the effective
Skyrme interaction energy density functional without the spin-orbit term.
The isospin asymmetry $\delta = (\rho_{n}-\rho_{p})/(\rho_{n}+\rho_{p})$
where $\rho$, $\rho_{n}$, and $\rho_{p}$ are the nucleon,
neutron, and proton densities, respectively.
In the present work, the parameter set IQ2 is used~\cite{Wang2005_MPLA20-2619}.
\label{modification:IQ2}

After integration, one obtains the local interaction potential energy:
\begin{eqnarray}
U_{\rm loc}&=&\frac{\alpha}{2}\sum_{i}\sum_{j\neq i}\frac{\rho_{ij}}{\rho_{0}}
              +\frac{\beta}{\gamma+1}\sum_{i}\left(\sum_{j\neq i}\frac{\rho_{ij}}{\rho_{0}}\right)^{\gamma}\nonumber\\
              &+&\frac{g_{\rm sur}}{2}\sum_{i}\sum_{j\neq i}f_{sij}\frac{\rho_{ij}}{\rho_{0}}
              +g_{\tau}\sum_{i}\left(\sum_{j\neq i}\frac{\rho_{ij}}{\rho_{0}}\right)^{\eta}\nonumber\\
              &+&\frac{C_{s}}{2}\sum_{i}\sum_{j\neq i}t_{i}t_{j}\frac{\rho_{ij}}{\rho_{0}}(1-\kappa_{s}f_{sij}),\label{eq1-6}
\end{eqnarray}
where
\begin{eqnarray}
\rho_{ij}&=&\frac{1}{(4\pi\sigma^{2}_{r})^{3/2}}
            \exp\left[-\frac{(\textbf{r}_i-\textbf{r}_{j})^{2}}{4\sigma^{2}_{r}}\right],\nonumber\\
f_{sij}&=&\frac{3}{2\sigma_{r}^{2}}
            \exp\left[-\left(\frac{\textbf{r}_i-\textbf{r}_{j}}{2\sigma^{2}_{r}}\right)^{2}\right],\label{eq1-7}
\end{eqnarray}
and $t_{i}=+1$ for protons and $-1$ for neutrons. The Coulomb energy is expressed as the
sum of the direct and the exchange contribution
\begin{eqnarray}
U_{\rm Coul}&=&\frac{1}{2}\int\int\rho_{p}(\textbf{r})\frac{e^{2}}{|\textbf{r}
         -\textbf{r}'|}\rho_{p}(\textbf{r}')d\textbf{r}'d\textbf{r}\nonumber\\
         &&\mbox{}
         + e^{2}\frac{3}{4}\left(\frac{3}{\pi}\right)^{1/3}\int \rho_{p}^{4/3}d\textbf{r}.
\label{eq1-8}
\end{eqnarray}

For low energy nuclear reactions, the Pauli principle plays an important role~\cite{
Feldmeier2000_RMP72-655,
Sargsyan2014_PRA90-022123}.
In the ImQMD model, by using the phase space occupation constraint
method~\cite{Papa2001_PRC64-024612}, the fermionic properties of nucleons
is approximately taken into account.

Many ImQMD simulations are made and each of them is called one event.
We examine the average properties of these events and the deviation of each event from the average.
For each event, to prepare the initial nuclei for the projectile and the target,
the position and momentum of each nucleon are randomly
given under certain macroscopic conditions, such as the binding energy and the radius.
Numerical details can be found in Refs.~\cite{Tian2008_PRC77-064603, Zhao2009_PRC79-024614}.

\subsection{\label{sec:II-B}Macroscopic Reduction Procedure for ImQMD model}

Since we focus on symmetric fusion reactions with the impact parameter
equal to zero, we can introduce a separation plane at the center of mass (CoM)
of the whole system, and divide the whole system into the left- (projectile-like) and
right- (target-like) half parts.
The relative motion between two CoMs of the left and right parts is
chosen to be the relevant degree of freedom to be described by the Langevin equation.

The one-dimensional generalized Langevin equation with memory effects
is given as~\cite{Gardiner1991, Mori1965_PTP33-423, Sakata2011_PTP125-359,
Frobrich1998_PR292-131}
\begin{eqnarray}
\frac{dP(t)}{dt}
    =-\int_{-\infty}^t \gamma (t-t')P(t')dt'
     + {\delta F}(t)
     -\frac{dU(R)}{dR},\label{eq1-9}
\end{eqnarray}
where $P(t)$ is the relative momentum between two parts, $\delta F(t)$ is the
random force felt by either part, and $U(R)$ is the collective potential between two parts.
The first term on the right hand side represents the retarded friction force.

Based on this stochastic equation, in the ImQMD simulations, the mean value of
the relative momentum $\langle P\rangle_{{R}}$ between two CoMs at a given
${R}$ is defined as
\begin{equation}
 \langle P \rangle_{{R}} \equiv
 \frac{1}{n} \sum_{i=1}^n \left. P_i(t_i) \right|_{\{ t_i|{R}_i(t_i)={R}\}},
\label{eq1-10}
\end{equation}
where $P_i(t)$ and ${R}_i(t)$ are the momentum and position of the $i$-th
event at time $t$. The time $t_i$ for the $i$-th event is chosen in such a
way that the relative distance takes a given value ${R}$, i.e., ${R}_i(t_i)={R}$.
In the present work, various microscopic variables are discussed as a function
of relative distance $R$ instead of time $t$, because $R$ characterizes how
near the two nuclei locate.
In Eq.~(\ref{eq1-10}) and hereafter,
$\langle {Q} \rangle$ denotes an average of ${Q}$ over all events.
When it does not cause any confusion, we hereafter use the same notation
$Q(R)$ for the single event $Q_i(R)$ as well as for the average
$\langle {Q}(R) \rangle =\sum_{i=1}^n Q_i(R)/n$ otherwise mentioned.

A collective potential for the relative motion is defined as,
\begin{eqnarray}
 U({R}) = E_{\mathrm{tot}}({R})
            - E_{\mathrm{left}}({R}) - E_{\mathrm{right}}({R}),
~\label{eq1-11}
\end{eqnarray}
where $E_{\mathrm{tot}}({R})$, $E_{\mathrm{left}}({R})$,
and $E_{\mathrm{right}}({R})$ represent the
energy of the total system and those of the left and right parts, respectively.
Each of them consists of the kinetic energy, the nuclear and the Coulomb
potential energies, and numerical results of $U({R})$ are shown in
Figs. \ref{fig:fluc-diss} and \ref{fig:3energy}(b).
The DC- and DD-TDHF have been applied to extract the collective potentials
between two nuclei~\cite{Umar2006_PRC74-061601R,Washiyama2008_PRC78-024610}
which gave similar features as those obtained in the present work and in
other QMD simulations~\cite{Jiang2010_PRC81-044602, Zanganeh2012_PRC85-034601}.

\label{modification:mu}
With the collective potential and the momentum, we define the collective energy as
\begin{eqnarray}
E_{\mathrm{coll}}({R})&=&T_{\rm coll}(R)+U({R}),\nonumber\\
T_{\rm coll}(R)&=&\frac{\langle P \rangle_R^2}{2\mu },
\label{eq1-12}
\end{eqnarray}
where $\mu$ is the reduced mass of the system.
After the two nuclei contact each other, $\mu$ becomes dependent on
the relative distance $R$ and $E_\mathrm{c.m.}$~\cite{Washiyama2008_PRC78-024610,
Washiyama2009_PRC79-024609, Zhao2009_PRC79-024614}.
As our discussion is limited to regions of $R$ and $E_\mathrm{c.m.}$
where $\mu$ does not change largely, in this work we take it to be a constant.

In the ImQMD simulations, the random force for the $i$-th event is defined as
\begin{eqnarray}
 {\delta F_{i}(R)}  &\equiv &  F^{i}(R) - \langle F(R) \rangle,\\
 F^{i}(R) &\equiv & \sum_{j=1}^{A} \left. f^{j}_i(t)\right|_{\{t|R(t)=R\}},
 \label{eq:14}
\\
 \langle F(R) \rangle &\equiv & \frac{1}{n} \sum_{i=1}^{n} F^{i}(R),
 \label{eq1-13}
\end{eqnarray}
where $f^{j}_{i}(t)$ denotes, in the $i$-th event, the force acting on
the $j$-th nucleon due to all other nucleons in the two fusing nuclei,
$A$ is the number of nucleons contained
in the left (right) part, and $n$ is the total number of events.

The fluctuation-dissipation relation links the energy dissipation of the
collective energy $E_{\rm coll}(R)$ to the fluctuation force originated from
the microscopic dynamics, both of which play decisive roles in the
macroscopic description of dissipation phenomena.
In the ImQMD simulations, the strength of the fluctuation is characterized by
\begin{equation}
 \langle {\delta F}({R}){\delta F}({R})\rangle
 \equiv
 \frac{1}{n} \sum_{i=1}^n  {\delta F}_i(R){\delta F}_i(R).
 \label{eq1-14}
\end{equation}

Assuming the work done by the collective motion against the friction force
is completely converted into the intrinsic energy $E_{\rm intr}(R)$, we get
a relation
\begin{equation}
 E_{\rm intr}({R}) \equiv  E_{\rm c.m.} - E_{\rm coll}({R}),\label{eq1-15}
\end{equation}
where $E_{\mathrm{c.m.}}$ means the initial bombarding energy.
\label{modification:gamma}
With the aid of the Rayleigh's dissipation function~\cite{Goldstein1959, Koonin1976_PRC13-209}
defined as
\begin{equation}
{\cal F}\equiv \mu\gamma(R)(dR/dt)^2, \label{eq1-16}
\end{equation}
where $\gamma(R)$ expresses the friction parameter
without considering the non-Markovian effects, the rate of
energy loss from the collective motion (which is equivalent to the energy gain
of the intrinsic system under the above assumption) is expressed
as $dE_{\rm intr}= {\cal F}dt$. From these relations, one gets
\begin{eqnarray}
\frac{dE_{\rm intr}(R)}{dR}=\mu\gamma(R)\frac{dR}{dt}=\gamma(R)P.\label{eq1-17}
\end{eqnarray}
The friction parameter $\gamma_{0}$ under the Markovian approximation is
expressed as
\begin{eqnarray}
 \gamma_0({R}) \equiv \frac{ \langle F_{\mathrm{fric}}({R})\rangle}{\langle P\rangle _{{R}}}, \quad
 F_{\mathrm{fric}}({R}) \equiv  \frac{dE_{\mathrm{intr}}({R})}{d{R}}.
~\label{eq1-18}
\end{eqnarray}
It should be noticed that the $R$-dependence of the friction parameter
$\gamma(R)$ contains a sensitive information on the dynamics of energy
transfer from the collective motion to the intrinsic degrees of freedom,
because it depends on the derivative of the collective potential $U(R)$
with respect to $R$.

\section{\label{sec:energy_dep}Incident Energy Dependence of Fluctuation and Dissipation}

\subsection{\label{sec:random}Random force in the ImQMD simulations}

\begin{figure*}[h]
\begin{centering}
\includegraphics[width=1.25\columnwidth]{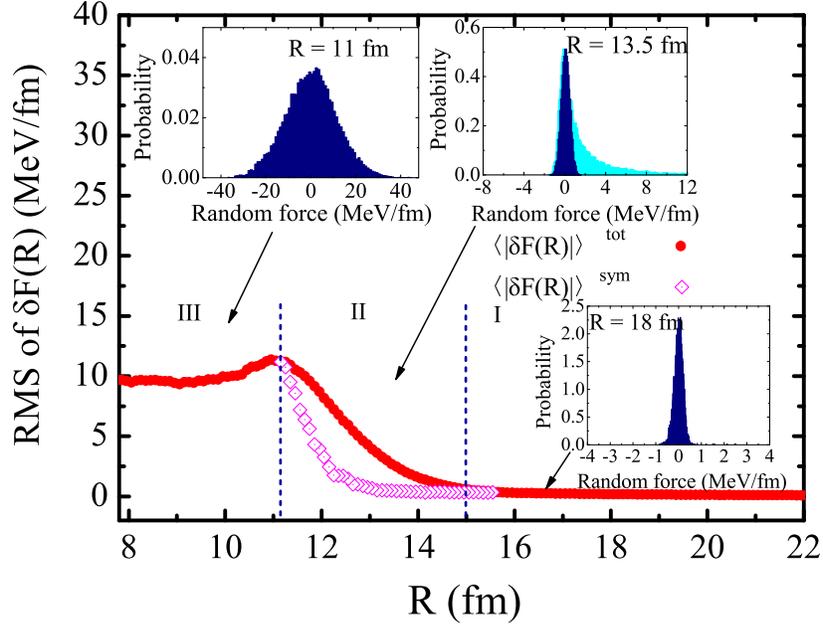}
\par\end{centering}
\caption{\label{fig:width}(Color online)
Root mean square of the random force distribution as a function of relative distance $R$.
Each inset shows a typical distribution in regions I, II, and III respectively.
In the inset corresponding to a typical distribution in region II,
the distribution is shifted so that the symmetric Gaussian distribution centers
at $\delta F = 0$, but the mean value of the random force is still zero.
The pink diamond is the root mean square of the random force 
after eliminating events in the asymmetric tail as shown
by the inset in region II.
}
\end{figure*}

\label{modification:fig1}
Figure \ref{fig:width} shows the width of the random force,
which is defined as the root mean square of $\delta F(R)$, i.e.,
the square root value of the strength of fluctuation defined in Eq. (\ref{eq1-14}).
In the case of $E_{\rm c.m.}=195$ MeV, the width of the random force turns
out to be a smooth function of relative distance $R$, reaching a peak at $R \cong 11$ fm
which locates slightly inside of the Coulomb barrier and then levels
out after a small down slope. At $R=30$ fm where we start our numerical
simulation, the fluctuation comes from the randomness of the position
and momentum of each particle at the initial time when each event is
initialized under the macroscopic conditions.
This fluctuation propagates following the equation of motion (\ref{eq1-3}).
Since the scattering angles of the nucleon-nucleon
collisions are randomly chosen in such a way that the distribution of
the scattering angles of all collisions agree with the measured angular
distribution for elastic and inelastic collisions in the QMD model
\cite{Aichelin1991_PR202-233},
the randomness coming from the scattering may also contribute to the fluctuation.
According to the shape of the distribution of the random force, we divide
the whole process into three regions \cite{Wen2013_PRL111-012501}.
As is seen from Fig.~\ref{fig:width}, the width is narrow and stays
unchanged at 15 fm $\lesssim R \lesssim$ 30 fm (region I), indicating the
stability of ImQMD simulations.

In region II (11 fm $\lesssim R \lesssim$ 15 fm) where the two nuclei are going
to merge into one nucleus, one may see that the width has a growing asymmetric
component in addition to the symmetric Gaussian.
It turns out \cite{Wen2013_PRL111-012501} that the asymmetric distribution
is caused by a small number of events where a few nucleons are exchanged
between two nuclei; these nucleons play a role in opening a {\it window}.
In the majority of events which contribute to the main part of the Gaussian
distribution, all nucleons well split into two separated groups, expressing
a projectile nucleus and a target nucleus and keeping their initial stable mean-fields.
As can be seen in Fig.~\ref{fig:width},
after eliminating those events where a few nucleons are exchanged between two nuclei, 
one gets a Gaussian distribution for the random force (the dark blue part
in the inset corresponding to $R=13.5$ fm)
and a smaller root mean square of the random force displayed
by the pink diamond in region II.

\label{modification:refs}
After this merging stage, we may see that the width of the random force in
region III has again a Gaussian shape which is, however, two order of magnitude wider
than that in region I.
The main origin of this enlargement is caused by the above discussed
transferred nucleons that feel the nuclear force from the other nucleus.
It is an open question
whether this factor is related to the formation of a neck or not
\cite{Siwek-Wilczynska2012_PRC86-014611,
Boilley2011_PRC84-054608,
Liang2012_EPJA48-133,
Zhu2013_CPL30-082401,
Zagrebaev2012_PRC85-014608,
Shen2002_PRC66-061602R,
Aritomo2012_PRC85-044614,
Liu2013_PRC87-034616,
Adamian1997_NPA619-241, Adamian2000_NPA671-233, Diaz-Torres2000_PLB481-228}.

\begin{figure*}
\begin{centering}
\includegraphics[width=1.60\columnwidth]{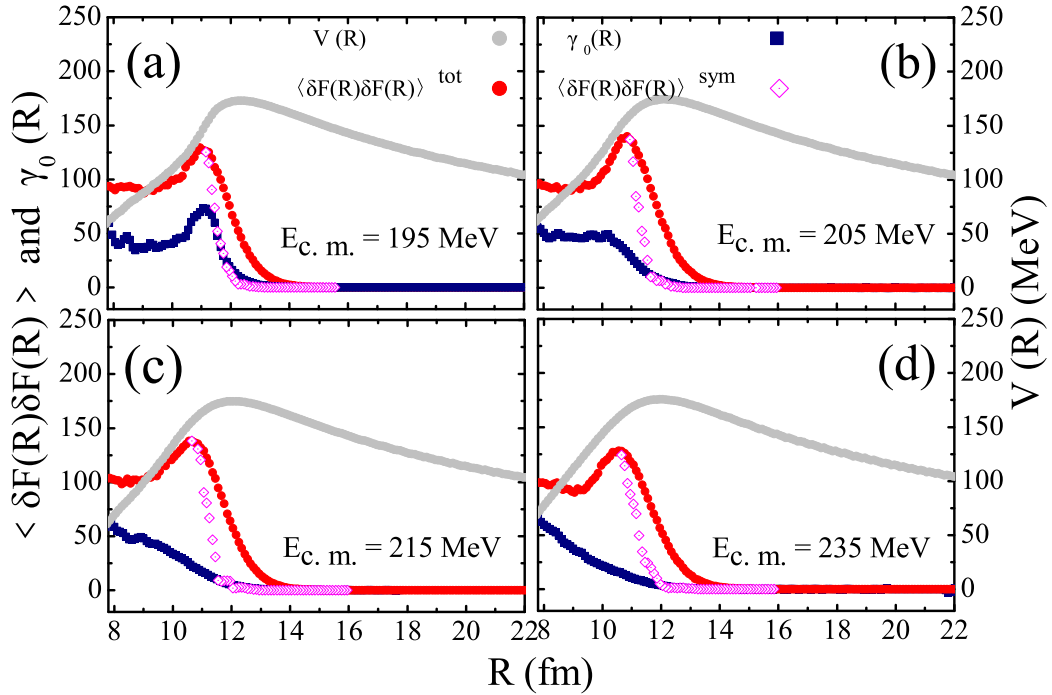}
\par\end{centering}
\caption{\label{fig:fluc-diss}(Color online)
The strength of random force
$\langle {\delta F}({R}) {\delta F}({R})\rangle^{\mathrm{tot}}$
[red dots, in (MeV/fm)$^{2}$]
and the friction coefficient $\gamma_0({R})$
(blue squares, in $0.001~c$/fm) at $E_{\rm c.m.}$ = 195 MeV, 205 MeV, 215 MeV,
and 235 MeV, respectively. The grey line shows the potential $U({R})$.
Pink diamonds represent $\langle {\delta F}
({R}){\delta F}({R})\rangle^{\mathrm{sym}}$ which is obtained by eliminating
events in the asymmetric tail.
}
\end{figure*}

Figure \ref{fig:fluc-diss} shows how the strength of random force and
the friction parameter $\gamma_0(R)$ defined in Eq.~(\ref{eq1-18})
change depending on the initial bombarding energy $E_{\rm c.m.}$.
One may see that the fluctuation $\langle {\delta F}({R}){\delta F}({R})\rangle$
and the friction coefficient $\gamma_0({R})$ at $E_{\rm c.m.}=$ 195 MeV show
similar shapes and their peaks locate at the same point.
Such a bumped shape in the friction parameter near the Coulomb barrier
energy is also observed in the DD-TDHF calculations~\cite{Washiyama2009_PRC79-024609,
Ayik2009_PRC79-054606}.

When $E_{\rm c.m.}$ increases, the friction parameter exhibits a sizable
energy dependence. Comparing the case of $E_{\rm c.m.}=$ 195 MeV with that
of $E_{\rm c.m.}=$ 205 MeV [Figs.~\ref{fig:fluc-diss}(a) and (b)], one may
recognize that the peak in the friction parameter disappears, while the
strength of the random force keeps its shape almost unchanged.
That is, a similarity between these two curves at energies just above the Coulomb barrier
gradually disappears as $E_{\rm c.m.}$ increases.
When $E_{\rm c.m.}$ increases more, the friction
parameter shows a monotonically increasing behavior [Figs.~\ref{fig:fluc-diss}(c) and (d)].
At the region near the barrier top $11$ fm$\lesssim R \lesssim 12$ fm, that is, the
initial stage of fusion just after the capture, the higher the incident
energy, the smaller the friction parameter,
while the situation becomes reverse when the two colliding nuclei
come closer with each other.

Note that in the ImQMD simulations, the nucleon collision effects
are included in addition to the one-body dissipation.
From the above discussions, it is seen that the ImQMD model
gives similar energy dependence and magnitude of the
friction parameter as those of the TDHF
calculations~\cite{Washiyama2009_PRC79-024609,Ayik2009_PRC79-054606,Umar2014_PRC89-034611},
indicating a dominance of one-body dissipation in the energy region under discussion.

\subsection{\label{sec:3energy}Incident Energy Dependences of Collective Kinetic,
Potential, and Intrinsic Energies}

Recently, it becomes clear that the collective potential
$U(R)$ exhibits rather sensitive energy dependence when $E_{\rm c.m.}$
takes near the Coulomb barrier energy \cite{Wu2004_HEPNP28-1317,
Washiyama2008_PRC78-024610, Umar2014_PRC89-034611}.
It has been discussed that this energy dependence might be related
to the dynamical change from adiabatic to sudden fusion process:
The two fusing nuclei have enough time to rearrange their
intrinsic structure when $E_{\rm c.m.}$ is just above the Coulomb barrier energy, and the
non-adiabatic effects gradually play a role as $E_{\rm c.m.}$ increases.
However, the above arguments are mainly based on the bulk information
like the density distribution of many nucleons, without referring
to any microscopic dynamics of individual nucleons.
Next we investigate the incident energy dependences of
the collective kinetic energy $T_\mathrm{coll}(R) \equiv E_\mathrm{coll}(R) - U(R)$,
the potential energy $U(R)$, and the intrinsic energy $E_\mathrm{intr}(R)$
with the microscopic ImQMD model.

\label{modification:fig3}
\begin{figure}
\begin{centering}
\includegraphics[width=0.90\columnwidth]{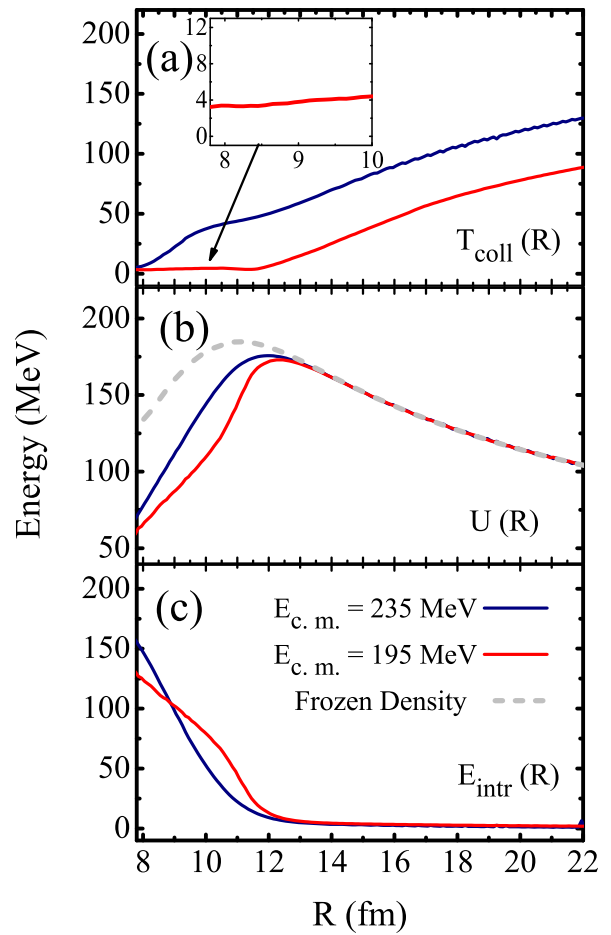}
\par\end{centering}
\caption{\label{fig:3energy}(Color online)
(a) Collective kinetic energy $T_{\rm coll}(R)=E_{\rm coll}(R)-U(R)$,
(b) potential energy $U(R)$, and
(c) intrinsic energy $E_\mathrm{intr}(R)$
as functions of relative distance $R$.
The red line represents results at $E_{\rm c.m.} = 195$ MeV and
the blue line shows results at $E_{\rm c.m.} = 195$ MeV.
In (a), the inset shows the kinetic energy from 8 fm to 10 fm at $E_{\rm c.m.} = 195$ MeV.
In (b), the light grey line shows the potential energy calculated under the frozen approximation.
}
\end{figure}

Figure \ref{fig:3energy}(a) shows the $R$-dependence of collective
kinetic energy $T_{\rm coll}(R)$ as a function of relative
distance at $E_{\rm c.m.} = 195$ MeV and $E_{\rm c.m.} = 235$ MeV.
In the case of $E_{\rm c.m.} = 195$ MeV, the main part of the collective
kinetic energy has already changed into a form of collective potential energy just
after the collective motion overcomes the Coulomb barrier at $R\cong 12.5$ fm.
Since there remains about 3 MeV rather constant collective kinetic energy
in 8 fm $\leq R \lesssim$ 11.5 fm as is shown in the inset of
Fig. \ref{fig:3energy}(a), the fusing process takes place very slowly.
In the case of $E_{\rm c.m.} = 235$ MeV, after passing through the barrier,
the collective kinetic energy takes almost ten times larger value than
that in the case of $E_{\rm c.m.} = 195$ MeV, and shows continuous down
slope without reaching a constant value.
Here, it should be mentioned that the dissipated energy (equivalent to $E_{\rm intr}$)
at $E_{\rm c.m.} = 235$ MeV is much smaller than that at $E_{\rm c.m.} = 195$ MeV
in the region 9 fm $\lesssim R \lesssim$ 11.5 fm as shown in Fig. \ref{fig:3energy}(c),
which is caused by the same reason why the friction parameter gets reduced as $E_{\rm c.m.}$
increases. This point will be clarified in Sec. \ref{sec.4}.

By comparing $E_{\rm intr}(R)$ in Fig. \ref{fig:3energy}(c) with $U(R)$ and $T_{\rm coll}(R)$
in Figs. \ref{fig:3energy}(a) and \ref{fig:3energy}(b), one finds that the
dissipated energy at $E_{\rm c.m.} = 235$ MeV
is around 30 MeV smaller than that at 195 MeV,
while $U(R)$ is around 30 MeV larger at $R\cong$ 10 fm.
This means that the friction force is not proportional but inversely
proportional to the momentum of the collective motion at the initial
stage of fusion process at $R\cong$ 10 fm.
In Fig.~\ref{fig:3energy}(b),
the grey line shows the potential energy calculated under the frozen approximation,
i.e., the potential calculated by fixing the density distribution of the projectile
and target and making them overlapped at a given $R$.
The barrier height is about 182 MeV.
The deviation of $U(R)$ from this potential energy under the frozen approximation
could be attributed to the rearrangement effect of all the intrinsic degrees of
freedom, which will be discussed in the next section.

\section{\label{sec:micro_dyn}Microscopic Dynamics of Energy Dissipation}~\label{sec.4}

In the previous Section, we have discussed $E_{\rm c.m.}$-dependence of
the nucleus-nucleus potential $U(R)$ as well as that of the friction
parameter $\gamma_0(R)$. 
Since the significant $E_{\rm c.m.}$-dependence occurs near
the Coulomb barrier energy where the applicability of the adiabatic
approximation is gradually replaced by that of the sudden approximation
when $E_{\rm c.m.}$ increases, let us
introduce new macroscopic quantities and further develop
the macroscopic reduction procedure.

\label{modification:W}
\subsection{\label{sec:IV-A}New Quantities Based on Macroscopic Reduction Procedure}~\label{sec4.1}

In this Subsection, we discuss how the collective energy is transferred
into the intrinsic one by dividing the dissipation into two processes.
The first one is caused by the nucleon transfer which takes place frequently
between the left and right parts of the whole system,
and the amount of energy transferred from the relative kinetic
energy to the intrinsic one up to a point $R$ is given by $T_{\rm diss}(R)$
in Eq.~(\ref{eqA-2}) in Appendix~\ref{app1}.

To discuss the second process of dissipation, we pay
attention to $F_{\rm intr}^{i}(R)$ for the $i$-th event which expresses the
difference between the force obtained by differentiating the nucleus-nucleus
potential $U(R)$ in Eq. (\ref{eq1-11}) and that obtained by summing all the nucleon-nucleon
forces in Eq. (\ref{eq:14}). $F_{\rm intr}^{i}(R)$ is defined as,
\begin{eqnarray}
&&\qquad F_{\rm intr}^i(R)\equiv F_{\rm tot}^i(R)-F^i_{\rm coll}(R),\cr
&&F_{\rm tot}^i(R)\equiv -\frac{\partial U^i(R)}{\partial R}, \quad F^i_{\rm coll}(R)\equiv F^i(R), \label{eq1-19}
\end{eqnarray}
where $U^i(R)$ and $F^i(R)$ for the $i$-th event are given in Eqs. (\ref{eq1-11}) and (\ref{eq:14}), respectively.
For the sake of convenience, $F^i_{\rm coll}(R)$ and $F_{\rm tot}^i(R)$ will be called as the
collective force and the total nucleus-nucleus force, respectively.
Corresponding to the collective force $F^i_{\rm coll}(R)$, 
one may introduce another potential $U^i_{\rm coll}(R)$ defined by
\begin{eqnarray}
U_{\rm coll}^i(R)\equiv -\int_{\infty}^R F^i_{\rm coll}(R')dR',\quad U^i_{\rm coll}(\infty)=0,\label{eq1-20}
\end{eqnarray}
which expresses the work done by the collective system against $F^i_{\rm coll}(R)$ up to the point $R$.
Using $U_{\rm coll}(R)$ together with $U(R)$ in Eq. (\ref{eq1-11}), in addition to
the difference in Eq. (\ref{eq1-19}), one may further introduce a deference as
\begin{eqnarray}
W(R)&=&-\int_{\infty}^{R}dR' (F_{\rm tot}^{i} - F_{\rm coll}^{i}) \nonumber\\
 &=& -\int_{\infty}^{R} dR' F_{\rm intr}(R').\label{eq1-21}
\end{eqnarray}

As is well known, the reduction of many-body dynamics onto a one-dimensional
collective space inevitably introduces additional effects coming from
the intrinsic degrees of freedom, besides the genuine effect within the collective space.
That is, $F_{\rm intr}^i(R)$ and $W(R)$ originally acting in the intrinsic space
are expressed as the additional force and additional potential in
the collective space because of eliminating the intrinsic degrees of freedom,
and of changing the time-scale into the collective space-scale introduced
in Eq. (\ref{eq1-10}).
More generally, it is shown analytically~\cite{Sakata2014_in-prep} that
the dynamics in the collective system and that in the intrinsic system are
distinguishable 
when one applies the macroscopic reduction procedure on the ImQMD simulations.

The additional potential $W(R)$ is then regarded as the work done by
the intrinsic system to rearrange its state which has been disturbed by
the transferred nucleons.
In other words, $-W(R)$ represents the second process which changes the energy in the intrinsic system.
With the aid of $W(R)$ and $T_{\rm diss}(R)$, the total amount of energy transferred
from the relative motion to the intrinsic one is given by
\begin{eqnarray}
E_{\rm{diss}}(R)=T_{\rm diss}(R)-W(R),\label{eq1-25}
\end{eqnarray}
where the former is from the collective kinetic energy through
the nucleon transfer, and the latter is due to the rearrangement of the intrinsic system.

To test our calculation, we may pay attention to a new quantity defined through
\begin{eqnarray}
 K_{\rm diss}(R)& \equiv &E_{\rm c.m.}-\left\{ \frac{P^2}{2\mu}+U_{\rm coll}(R)\right\}, \label{eq1-22}
\end{eqnarray}
where $U_{\rm coll}(R)$ is defined in Eq. (\ref{eq1-20}). It should be noticed 
that there holds an energy conservation law within the collective degree of freedom as
\begin{eqnarray}
 E_{\rm c.m.}\cong T_{\rm diss}(R)+ \left\{ \frac{P^2}{2\mu}+U_{\rm coll}(R)\right\}, \label{eq1-23}
\end{eqnarray}
where $T_{\rm diss}(R)$ means the dissipated energy due to nucleon
exchange defined in Eq. (\ref{eqA-2}) as mentioned at the beginning of this Subsection,
provided that there holds the relation
\begin{eqnarray}
 T_{\rm diss}(R)\cong K_{\rm diss}(R). \label{eq1-24}
\end{eqnarray}
In the following, the validity of Eq. (\ref{eq1-24}) will be shown with numerical simulations.
By using Eq. (\ref{eq1-25}),
the approximate energy conservation law in Eq. (\ref{eq1-23}) is then expressed as
\begin{eqnarray}
E_{\rm c.m.}\cong E_{\rm diss}(R)+\frac{P^2}{2\mu}+U_{\rm coll}(R)+W(R).\label{eq1-26}
\end{eqnarray}
Since the relation
\begin{equation}
 E_{\rm intr}(R)\approx E_{\rm diss}(R) ,
\label{eq1-27}
\end{equation}
is satisfied [c.f. Fig.~\ref{disspation} and relevant discussions],
Eq.~(\ref{eq1-26}) is reduced to Eq. (\ref{eq1-15}) which expresses
an energy conservation of the total system.

\subsection{Two Microscopic Processes of Energy Dissipation}~\label{sec5.1}

\begin{figure}
\begin{centering}
\includegraphics[width=0.90\columnwidth]{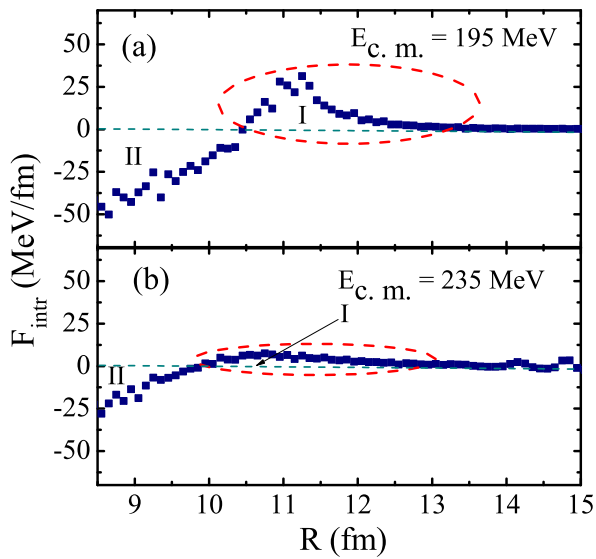}
\par\end{centering}
\caption{\label{fig6:zu3}(Color online)
$F_{\rm intr}(R)$, the difference between the total nucleus force
$F_{\rm tot}=-\partial U(R)/\partial R$ and the collective force $F_{\rm coll}(R)$
for the case of (a) $E_{\rm c.m.} = 195$ MeV and (b) $E_{\rm c.m.} = 235$ MeV.
``I'' and ``II'' in each subfigure correspond to regions I and II defined in Fig.~\ref{fig:width}.
}
\end{figure}

Making clear what information we get from the energy dependence of the
collective potential $U(R)$, we first examine $F_{\rm intr}(R)$.
It expresses what happens in the intrinsic system when many nucleons are exchanged
between two nuclei, and when two nuclei closely approach each other.
It also tells us how these effects subsequently develop inside of the intrinsic system,
which are usually considered as the rearrangement of the intrinsic system.

As is seen from Fig.~\ref{fig6:zu3}, our numerical simulation tells us that 
$F_{\rm intr}(R)$
takes positive value in the range of 10.5 fm$\lesssim R \lesssim$ 12.5 fm (10.0 fm
$\lesssim R \lesssim$ 12.5 fm) and negative value at $R\lesssim$ 10.5 fm
($R \lesssim $10 fm) in the case of $E_{\rm c.m.} = 195$ ($E_{\rm c.m.} = 235$) MeV.
It is also recognized from this figure that $F_{\rm intr}(R)$ in the case of
$E_{\rm c.m.} = 195$ MeV shows much stronger $R$-dependence than that of
$E_{\rm c.m.} = 235$ MeV.
A steep slope of $F_{\rm intr}(R)$ in the region 9 fm $\lesssim R\lesssim$ 11 fm
at $E_{\rm c.m.} = 195$ MeV indicates
that the rearrangement of the intrinsic system takes place more strongly than
that in the case of $E_{\rm c.m.} = 235$ MeV.

Since the intrinsic system is kept unchanged in the region $R >$ 12.0 fm
irrespective of $E_{\rm c.m.}$, which is expected from Fig.~\ref{fig:3energy}(c),
it is reasonable that $F_{\rm intr}(R)$ does not take any appreciable value in this region.
In the case of $E_{\rm c.m.} = 195$ MeV, the adiabatic approximation is
expected to be applicable because of a strong rearrangement of the intrinsic system.
In the case of $E_{\rm c.m.} = 235$ MeV, however, there appears more gentle slope
which indicates rather weak rearrangement effect in the intrinsic system.
In this case, the main parts of original nuclei invade into each other with
much shorter distance $R$ so that the frozen density and/or sudden approximation
may have more sense than the case with $E_{\rm c.m.}=$ 195 MeV.

\begin{figure}
\begin{centering}
\includegraphics[width=0.90\columnwidth]{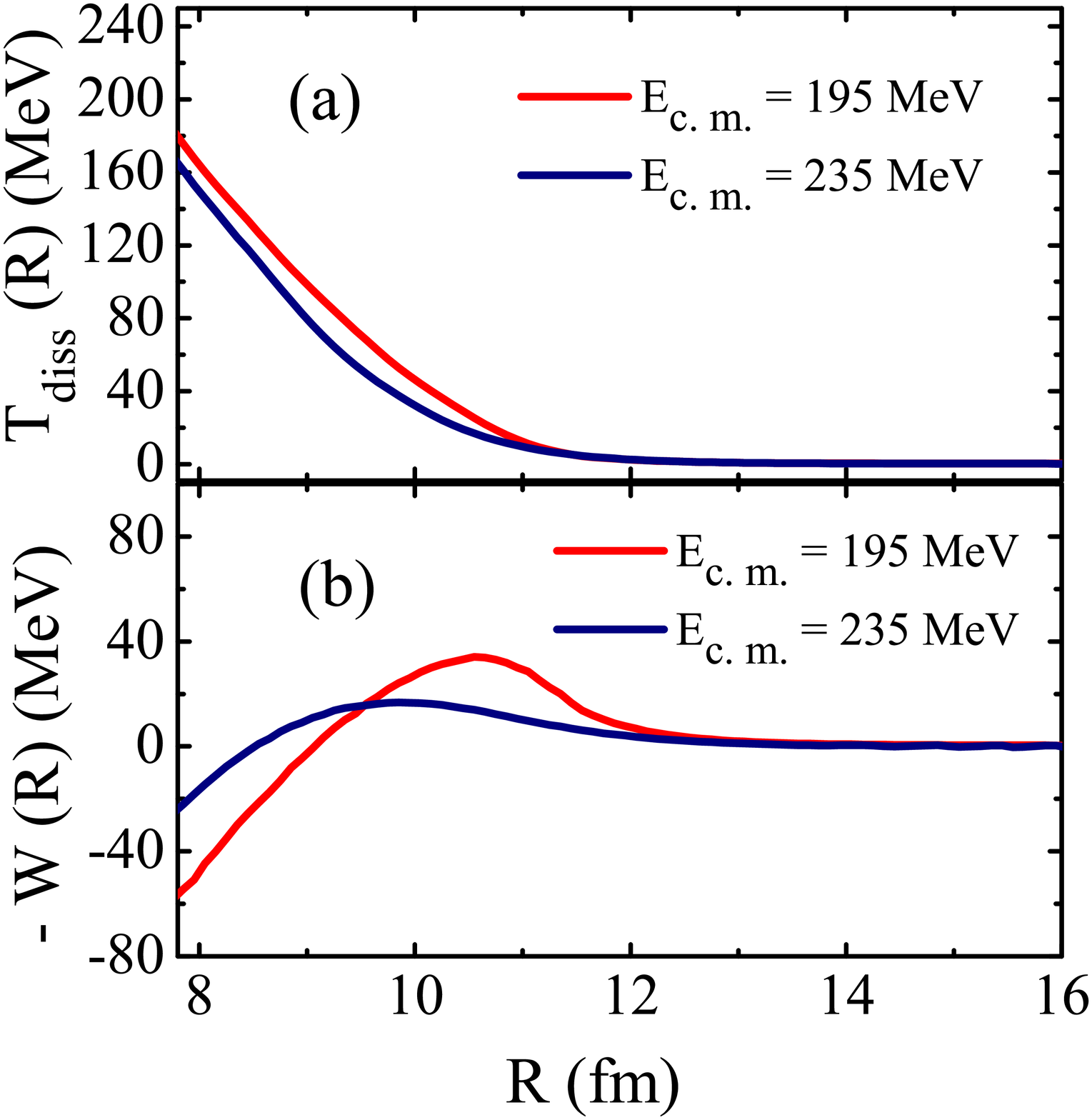}
\includegraphics[width=0.90\columnwidth]{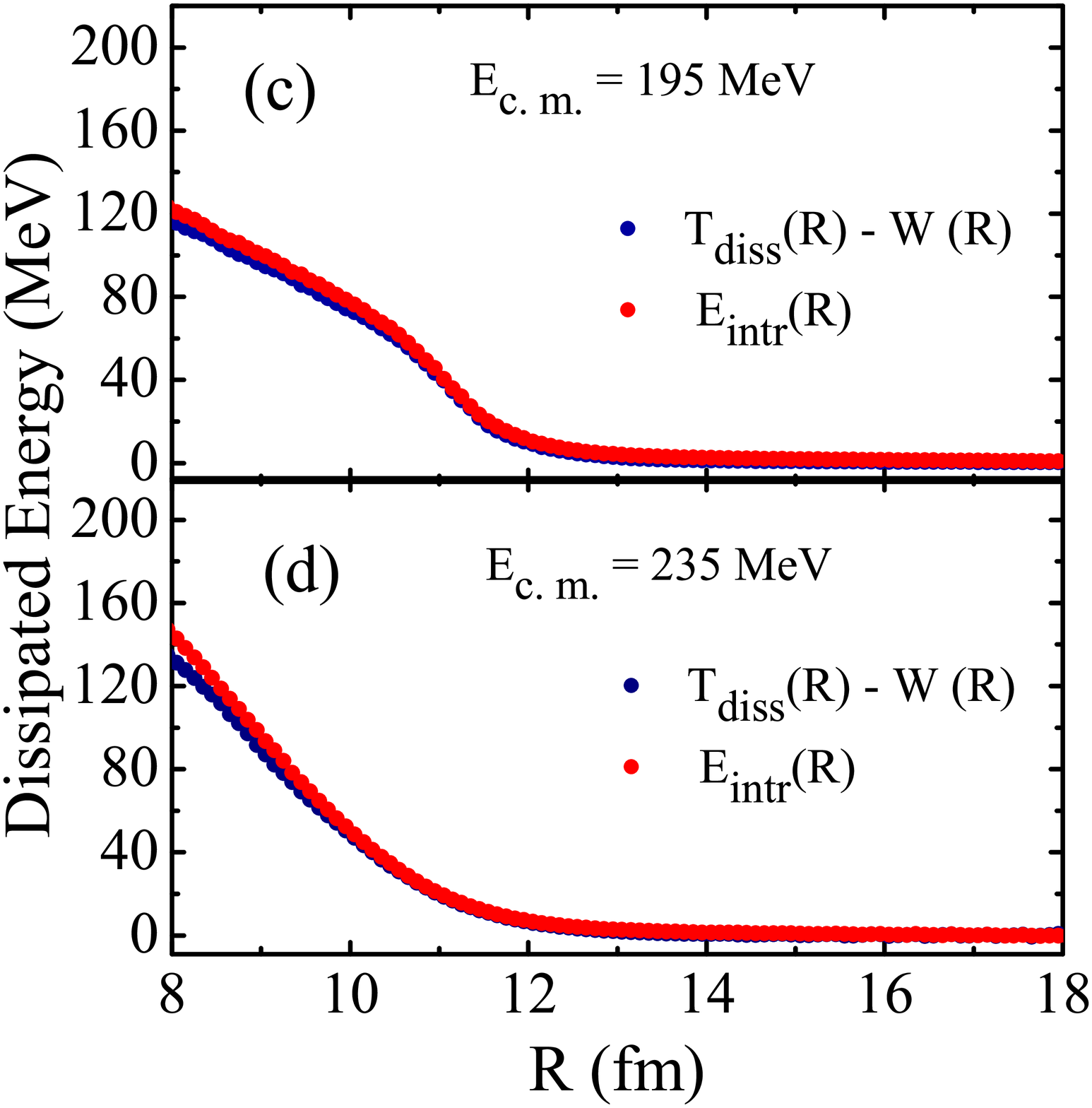}
\par\end{centering}
\caption{\label{disspation}(Color online)
Two microscopic sources of dissipation and its comparison with an
increase of intrinsic energy.
(a) $T_{\rm diss}(R)$ is from nucleon exchange process.
(b) $W(R)$ is due to the rearrangement effects in the intrinsic system.
(c) and (d) show the difference between two processes
$E_{\rm diss}=T_{\rm diss}(R)-W(R)$ compared to the intrinsic energy $E_{\rm intr}(R)$
at $E_{\rm c.m.} = 195$ MeV and $E_{\rm c.m.} = 235$ MeV respectively.
}
\end{figure}

Let us discuss $W(R)$ which is the integrated value of the difference
$F_{\rm intr}(R)$. Equation (\ref{eq1-25}) tells us that the
increase of $E_{\rm diss}(R)$ comes from two sources, i.e., $T_{\rm diss}(R)$ and $W(R)$.
As is discussed in Appendix~\ref{app1}, the former comes from the nucleon
transfer between two fusing nuclei, enhances the
dissipation, and might be related to the {\it window} effect.
On the other hand, the latter
is caused by the rearrangement process of the intrinsic system.
Namely, the disturbance due to the nucleon exchange, and the subsequent process
of forming a new state like an
excited mean-field in the intrinsic system play a role to first enhance the dissipation
and subsequently to suppress it.

In Fig.~\ref{disspation}, two microscopic sources of dissipation $T_{\rm diss}(R)$
and $W(R)$, and their combined effects $E_{\rm diss}(R)$ in comparison
with $E_{\rm intr}(R)$ defined in Eq.~(\ref{eq1-15}) are shown.
Figures \ref{disspation}(c) and \ref{disspation}(d) show that there well holds
the relation given in Eq.~(\ref{eq1-27}) irrespective of the initial bombarding energy.
From this figure, one may conclude that the energy transfer from the collective
motion to the intrinsic one
in the region 8 fm $\lesssim R\lesssim$ 11.5 fm can be understood microscopically
as follows.
Firstly, the nucleon exchange process always contribute largely to
the dissipation and the dissipated energy $T_{\rm diss}(R)$ can reach around 150 MeV.
Secondly, the rearrangement effect of intrinsic system
may either enhance or suppress the dissipation, depending on the relative
distance $R$.
Thirdly, the rearrangement effect is very sensitive to the incident energy.
For the case of $E_\mathrm{cm}$ = 195 MeV,
the rearrangement effect is very pronounced and $-60$ MeV $<W(R)<$ 40 MeV.
This is because that, the fusion takes place slowly and
there is much time for the whole system to rearrange nucleons.
In other words, the fusion process goes adiabatically.
However, for the case of $E_\mathrm{cm}$ = 235 MeV,
the rearrangement effect is less pronounced and $-20$ MeV $<W(R)<$ 20 MeV.
In this case, the fusion is mainly diabatic.
The adiabatic or diabatic characteristics in the fusion
will be discussed in more detail in next Subsection.

From the above numerical results, one may see that the nucleus-nucleus interaction
$F_{\rm tot}(R)$ is reduced to the simple summation of constituent forces
$F_{\rm coll}(R)$, provided there
are no internal correlations in the intrinsic system.
Namely $F_{\rm intr}(R)$ in Eq.~(\ref{eq1-19}) expresses
the amount of correlations inside the intrinsic system.

\begin{figure}
\begin{centering}
\includegraphics[width=0.90\columnwidth]{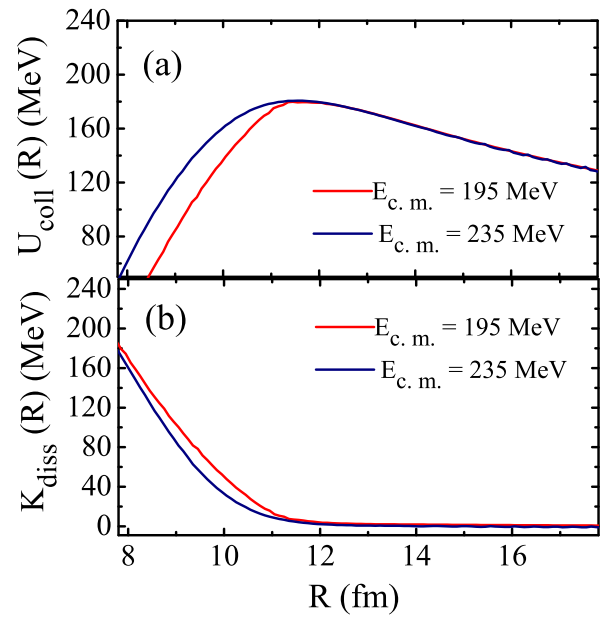}
\par\end{centering}
\caption{\label{fig:8}(Color online)
(a) Calculated value of $U_{\rm coll}(R)$ as a function $R$.
(b) Calculated value of $K_{\rm diss}(R)=E_{\rm c.m.}-P^2/2\mu-U_{\rm coll}(R)$
as a function $R$ defined in Eq.~(\ref{eq1-22}).
}
\end{figure}

At the end of this Subsection, we pay attention to another approximate way
of calculating $T_{\rm diss}(R)$ given by Eq.~(\ref{eqA-2}).
In Fig.~\ref{fig:8}, we show the calculated $U_{\rm coll}(R)$ and $K_{\rm diss}(R)$.
Comparing Fig.~\ref{fig:8}(b) with Fig.~\ref{disspation}(a), one can draw
a conclusion that there holds the relation in Eq. (\ref{eq1-24}).
That is, the dissipation energy due to the nucleon exchange
$T_{\rm diss}(R)$ is well reproduced by $K_{\rm diss}(R)$ defined in Eq. (\ref{eq1-22})
which only takes account of the potential $U_{\rm coll}(R)$ without
considering the internal correlation $W(R)$ of the intrinsic system.

\subsection{Adiabatic vs. Diabatic Processes }

In the previous Subsection, it is clearly shown that the fusion potential $U(R)$
observed in Fig.~\ref{fig:3energy}(c) can be divided into two different components
$U_{\rm coll}(R)$ and $W(R)$ which are depicted in
Figs.~\ref{fig:8}(a) and \ref{disspation}(b), respectively.
Since the energy dependence of $W(R)$ is discussed in the previous Subsection,
let us explore the energy dependence of $T_{\rm diss}(R)$ which gives
information on the incident energy dependence of $U_{\rm coll}(R)$.
From Figs.~\ref{disspation}(a) and \ref{fig:8}(b), the most important point
to be explored on the energy dependence of $T_{\rm diss}(R)$ and $K_{\rm diss}(R)$
is why the collective energy dissipation through the nucleon transfer occurs
more frequently at remote distance in the case of $E_{\rm c.m.} = 195$ MeV than
the case of $E_{\rm c.m.} = 235$ MeV.

\begin{figure}
\begin{centering}
\includegraphics[width=0.90\columnwidth]{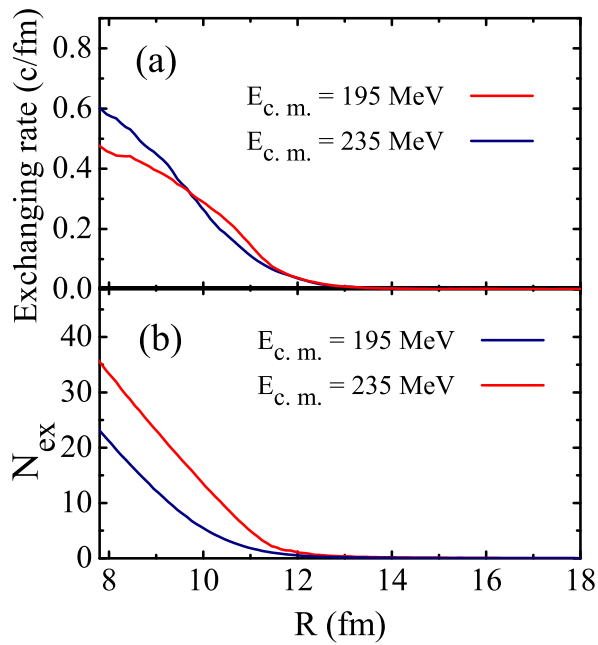}
\par\end{centering}
\caption{\label{fig9.exch}(Color online)
(a) Nucleon exchanging rate as a function of relative distance,
(b) Total number of nucleons which has ever transferred from
left (or right) part to the right (or left).
The red and blue lines represent two cases with $E_{\rm c.m.}=$ 195 and 235 MeV, respectively.}
\end{figure}

In Fig.~\ref{fig9.exch}(a), the nucleon exchange rate defined as the number
of nucleons which go through the separation plane per unit time is shown
as a function of $R$. In the region 9.5 fm $\lesssim R\lesssim$13 fm, there are almost
no appreciable incident energy dependence in the nucleon exchange rate, though it
becomes larger in the case of $E_{\rm c.m.} = 235$ MeV than the case of $E_{\rm c.m.} = 195$ MeV
for $R\lesssim$ 9.5 fm.
In Fig.~\ref{fig9.exch}(b), the total number of nucleons $N_{\rm ex}(R)$
which have ever transferred up to a distance $R$ is shown.
From Fig.~\ref{fig:3energy}(a), one may see that the kinetic energy of
relative motion around $R\approx$ 13 fm at $E_{\rm c.m.} = 235$ MeV is
more than 40 MeV larger than that at $E_{\rm c.m.} = 195$ MeV.
Namely, the velocity in the former case is a few times faster than
that in the case at $E_{\rm c.m.} = 195$ MeV.
From Figs.~\ref{fig:3energy}(a) and \ref{fig9.exch}(b),
one may recognize that, at $E_{\rm c.m.} = 235$ MeV,
the nuclei approach each other too fast to exchange comparable amount of
nucleons as in the case of $E_{\rm c.m.} = 195$ MeV.
In other words, the two nuclei at
$E_{\rm c.m.} = 235$ MeV keep much larger part of their original shapes
into a small relative distance $R$ than that at $E_{\rm c.m.} = 195$ MeV.
This mechanism explains the rather slow startup of the $T_{\rm diss}(R)$
and $K_{\rm diss}(R)$ in the case of $E_{\rm c.m.} = 235$ MeV observed
in Figs. \ref{disspation}(a) and \ref{fig:8}(b).
Since there holds the relation given in Eq.~(\ref{eq1-22}), the above discussed {\it diabatic}
aspect of nucleon transfer process and the large collective momentum in
the region 8 fm $\lesssim R\lesssim$ 11 fm for the case of $E_{\rm c.m.} = 235$ MeV
are two microscopic ingredients to determine the amount of work $U_{\rm coll}(R)$
done by the collective degrees of freedom.

\begin{figure*}
\begin{centering}
\includegraphics[width=1.50\columnwidth]{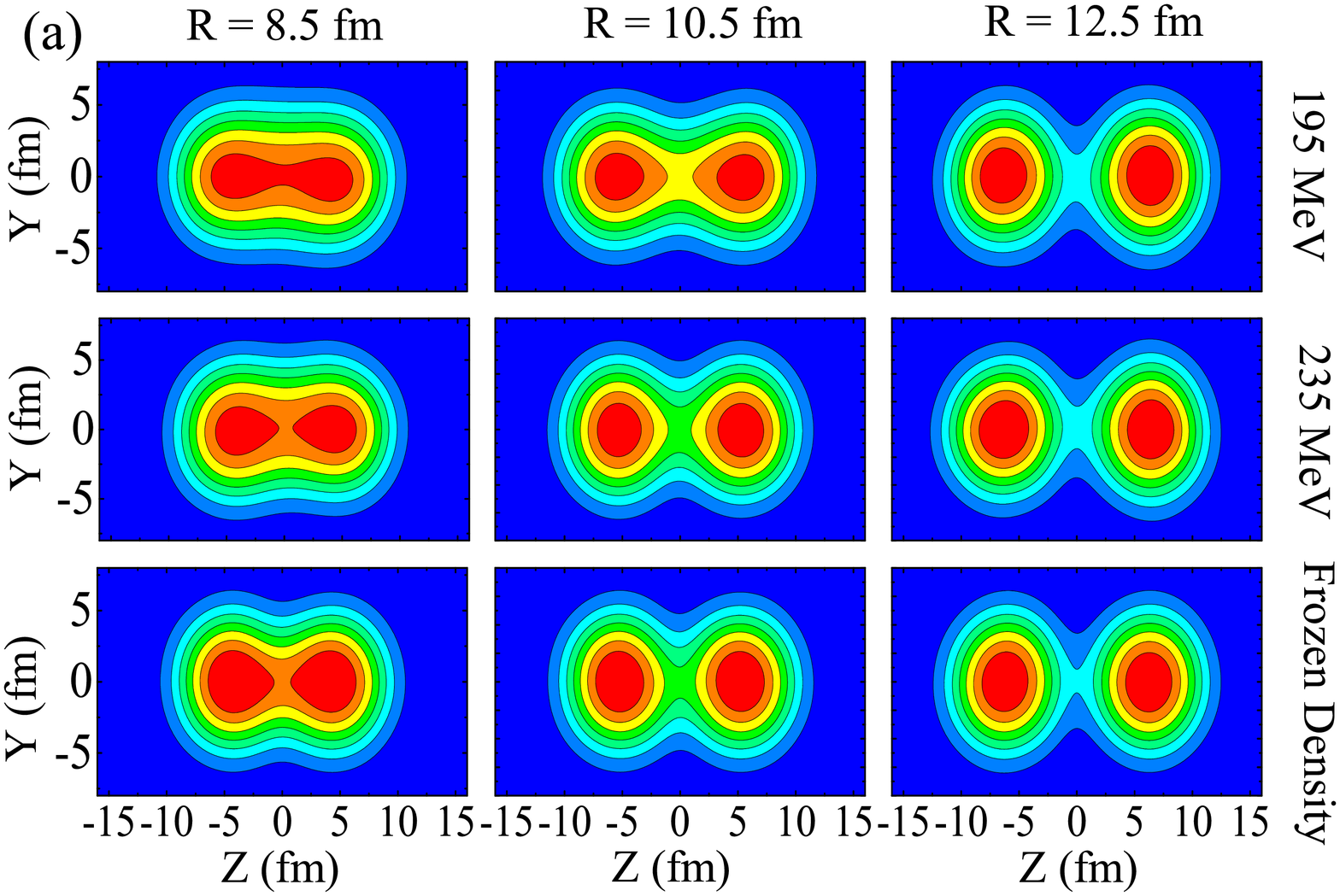}
\includegraphics[width=1.00\columnwidth]{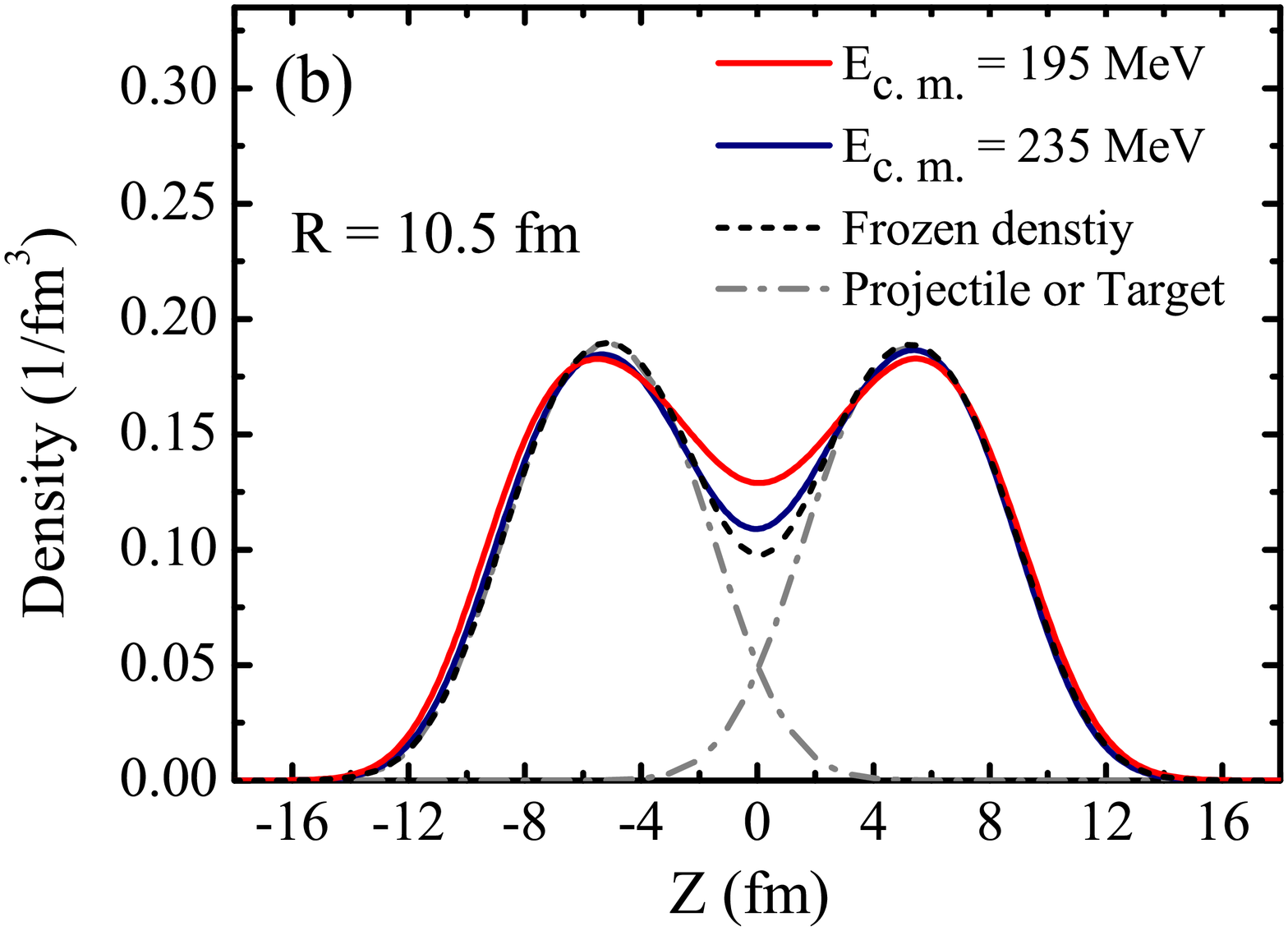}
\par\end{centering}
\caption{(Color online)
(a) Density profiles obtained in ImQMD for different relative
distances $R$ at $E_{\rm c.m.}$=195 MeV and 235 MeV and those with the
frozen density. The abscissa axis is the reaction axis $Z$ and the vertical axis is axis $Y$.
(b) Nucleon number density along the reaction axis $Z$ with
$X=Y=0$ at relative $R = 10.5$ fm obtained by the ImQMD model. The red line represents the
density at $E_{\rm c.m.} = 195$ MeV, the blue line at $E_{\rm c.m.} = 235$ MeV.
The dark grey dashed line is the density with frozen density,
the light grey dashed-dotted lines are densities for projectile and
target.~\label{fig:10density}}
\end{figure*}

In order to show whether or not our microscopic discussion based on the
role of $F_{\rm intr}(R)$ [and $W(R)$] is consistent with the understanding obtained
by the density profile, the density at $E_{\rm c.m.} = 195$ MeV and $E_{\rm c.m.} = 235$ MeV,
as well as that corresponding to the frozen density at different $R$ are depicted
in Fig.~\ref{fig:10density}(a).
The density profile under the frozen density approximation is obtained by
fixing the density distribution of the projectile and target and
making them overlapped at a given $R$ by hands.
An obvious difference in the density distribution among these three cases
is observed at the central part or the ``neck" region of the system.
This situation is displayed more clearly by the densities along the reaction
axis shown in Fig.~{\ref{fig:10density}}(b), which shows the nucleon number
densities along the reaction axis $Z$ with $X=Y=0$ at $R = 10.5$ fm.
From these figures, one may learn that the density at $E_{\rm c.m.} = 195$ MeV
in the neck region is obviously higher than the case at $E_{\rm c.m.} = 235$ MeV,
and the latter case gives a similar profile as that of the frozen density.
This lower density at the neck region in the latter case just corresponds to the
smaller number of transferred nucleons which explain the slower startup of the
transfer process $T_{\rm diss}(R)$ in the case of $E_{\rm c.m.} = 235$ MeV.

\subsection{Friction parameter }

We are now in the position to discuss the energy dependence of the friction
parameter $\gamma_0(R)$ shown in Fig. \ref{fig:fluc-diss}.
According to the macroscopic reduction procedure,
the friction parameter is derived in Eq.~(\ref{eq1-18}).
Consequently, the energy dependence of the friction parameter is directly related
to that of the nucleus-nucleus fusion potential which is shown in Fig.
\ref{fig:3energy}(b).
From Figs.~\ref{fig:3energy}(b) and (c), one may recognize that the $U(R)$ and
$E_{\rm intr}(R)$ at $E_{\rm c.m.} = 195$ MeV show an inflection point at
$R\approx$ 11.5 fm, while at $E_{\rm c.m.} = 235$ MeV such an inflection point does not exist.
This inflection point is responsible for the peak structure of the friction
parameter shown in Fig. \ref{fig:fluc-diss}(a).

\begin{figure}
\begin{centering}
\includegraphics[width=0.90\columnwidth]{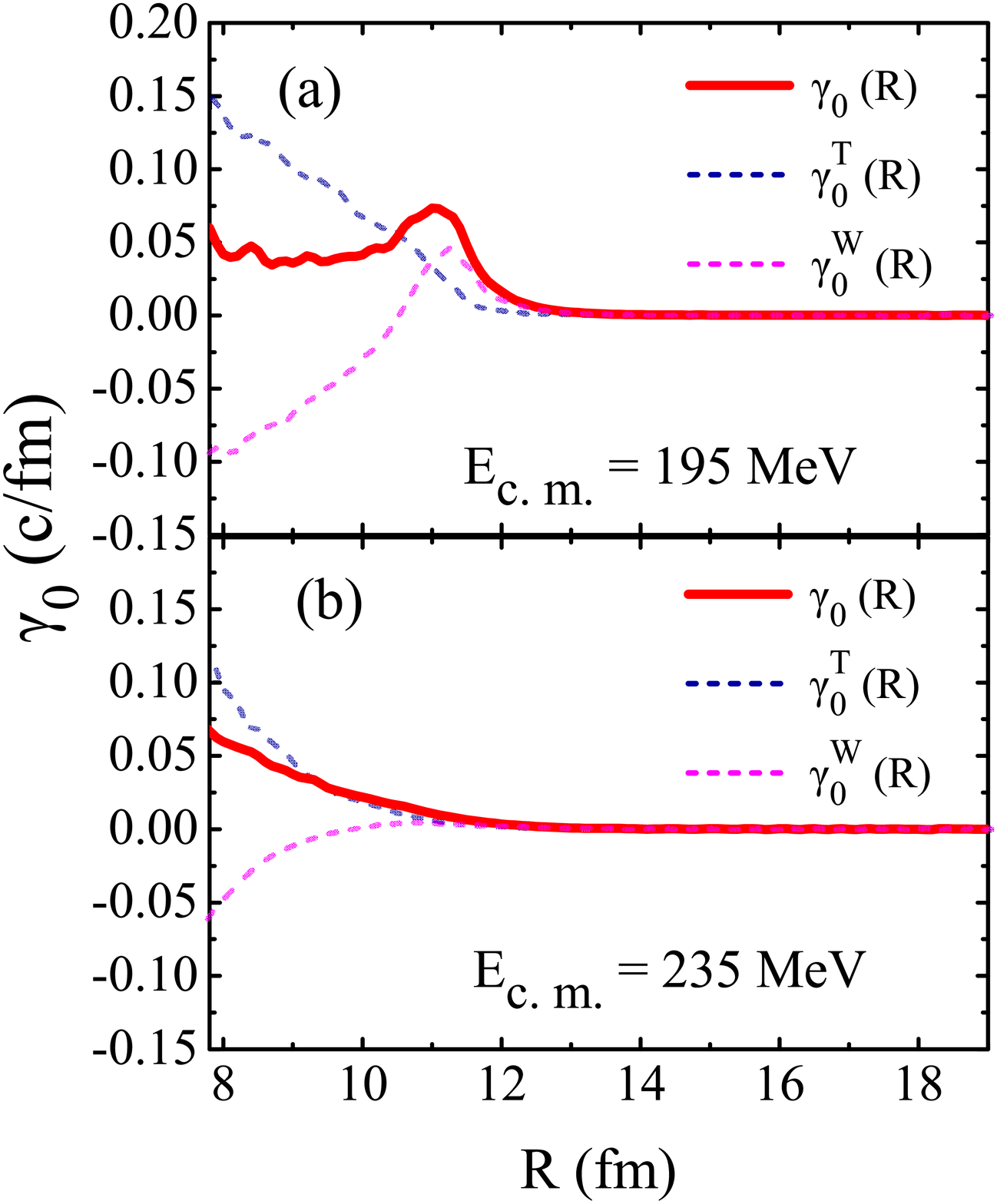}
\par\end{centering}
\caption{\label{fig11}(Color online)
$R$-dependence of friction parameter caused by two competitive microscopic processes at
(a) $E_{\rm c.m.} = 195$ MeV and (b) $E_{\rm c.m.} = 235$ MeV.
The blue dashed line shows effects of nucleon exchange.
The red dashed line shows effects of rearrangement in intrinsic system.
}
\end{figure}

Using $E_{\rm intr}(R)\simeq E_{\rm diss}(R)$, and from Eqs. (\ref{eq1-18})
and (\ref{eq1-25}), the friction parameter can also be expressed as
\begin{eqnarray}
\gamma_0(R)=\frac{1}{\langle P\rangle_R}\left\langle
                 \left(\frac{dT_{\rm diss}(R)}{dR}
                        -\frac{\partial W(R)}{\partial R} \right)\right\rangle.~\label{eq1-45}
\end{eqnarray}
which tells us that the friction parameter is determined by two microscopic
processes, i.e., the nucleon exchange and the rearrangement.
Since the $R$ dependence of $T_{\rm diss}(R)$ does not have the inflection
point irrespective of the incident energy in Fig.~\ref{disspation}(a), one
may see that the first derivative of $T_{\rm diss}(R)$ with respect to $R$
does not show a strong incident energy dependence.
On the other hand, $-W(R)$ shown in Fig. \ref{disspation}(b) has a strong
incident energy dependence.

In Fig.~\ref{fig11}, the $R$-dependence of the friction parameter is shown
together with the two components originating from two different microscopic processes.
The bumped structure of
the friction parameter in the case of $E_{\rm c.m.}$ = 195 MeV can be understood to
be a pronounced intrinsic effects at $R\simeq$ 11 fm
[c.f. $F_{\rm intr}(R)$ shown in Fig.~\ref{fig6:zu3}].
This bumped structure disappears rather quickly as $E_{\rm c.m.}$ increases,
because the rearrangement effects of the intrinsic system around
$R\cong 11.0$ fm become small, which is observed from Fig. \ref{fig6:zu3}.

At the end of this Subsection, some comments on the validity of the
fluctuation-dissipation relation should be addressed.
As is mentioned above, the $E_{\rm c.m.}$ dependence of the friction parameter is directly
related to the rearrangement effects in the intrinsic system expressed by $F_{\rm intr}(R)$.
On the other hand, the fluctuation $\delta F(R)$ is related to $F_{\rm coll}(R)$ which does
not depends on the rearrangement taking place in the
intrinsic system, and the shape change in the correlation function of the
fluctuation does not show an energy dependence as is seen from Fig. \ref{fig:fluc-diss}.
According to these different microscopic effects on the friction parameter and
on the fluctuation force, the fluctuation-dissipation relation realized at
$E_{\rm c.m.} = 195$ MeV breaks down when the incident energy increases.

\section{\label{sec:summary}Concluding Remarks}

In this paper, we have systematically studied the incident energy dependence
of the nucleus-nucleus potential, that of the friction parameter, as well as that
of the random force in heavy-ion fusion reactions, by applying the
macroscopic reduction procedure based on the ImQMD numerical simulation.
Making the discussion clear, we focus our attention on the head-on fusion
reaction of symmetric $^{90}$Zr+$^{90}$Zr system at above
the Coulomb barrier energies and pay special attention to the
fluctuation-dissipation relation.

It should be mentioned that our discussion is based on a separation of
the total nucleus-nucleus force $F_{\rm tot}(R)$ into the collective part
$F_{\rm coll}(R)$ and the intrinsic part $F_{\rm intr}(R)$.
Although the latter also depends on the relative distance $R$ rather than
the intrinsic coordinates, one may regard $W(R)$ as the work performed by
the intrinsic system against $F_{\rm intr}(R)$.
A theoretical justification of this point will be given in Ref.~\cite{Sakata2014_in-prep},
where a reduction of the many-body dynamics onto one-dimensional collective space is discussed.
In the present paper, it is numerically shown that there holds the relation
given in Eq.~(\ref{eq1-24}), i.e., $T_{\rm diss}(R)\approx K_{\rm diss}(R)$
which gives another justification for the above separation.
By exploiting the procedure appropriate for the ImQMD 
numerical
simulations, the first step toward the dynamics of emergence about how the macroscopic
fusion dynamics comes out as a result of the microscopic nucleonic motion is clarified.

We note that more study should be performed with non-zero impact parameters
and for different reaction systems, particularly for asymmetric reactions.
The spin-orbit coupling has been shown to be very important in
low-energy heavy-ion fusion reactions \cite{Umar1986_PRL56-2793,Dai2014_PRC}.
Furthermore, it was also found that the negative shell correction energies
lower potential barriers \cite{Zhu2013_NPA915-90}.
Therefore, more dissipations are expected if these effects are included
in the ImQMD simulations.
Finally, detailed investigations should also be made about the influence
of two-body collisions on the dissipation process in fusion reactions
at energies around the Coulomb barrier.

\appendix

\section{Derivation of $T_{\rm diss}(R)$}~\label{app1}

\begin{figure}[h]
\begin{centering}
\includegraphics[width=0.75\columnwidth]{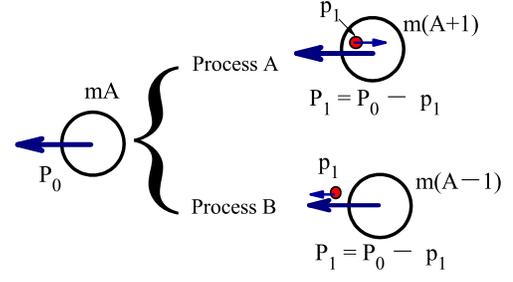}
\par\end{centering}
\caption{\label{transfer}(Color online)
The basic processes of nucleon exchange between two fusing nuclei.
}
\end{figure}

Let us discuss the nucleon exchange process between two nuclei with mass
$M=Am$ where $m$ denotes the nucleon mass and $A$ the number of nucleons in the nucleus.
Since we are interested in head-on symmetric heavy-ion collisions, we
focus on the one-dimensional case depicted in Fig. \ref{transfer} in which
only the right part of the system is shown.

At certain time $t_0$, the right part with mass $mA$ is assumed to move
toward the left part with the momentum $P_0$, and the kinetic energy $T_R^{(0)}=P_0^2/2mA$.
In the fusion reaction, there occur many nucleon transfers between two
nuclei which are characterized
by two basic processes depicted in Fig.~\ref{transfer}.
The one denoted as Process A is to get an additional nucleon with a momentum
$p_1$ heading to the right, which is originally from the left part of the system.
The other shown as Process B is to lose one nucleon with a momentum $p_1$
heading to the left. After the $i$-th nucleon transfer process, the kinetic
energy of the right part denoted as $T_R^{(i)}$ is expressed as
\begin{eqnarray}
&&T_R^{(i)}=\frac{P_i^2}{2mA_i},\quad  P_i=P_{i-1}-p_i,\quad A_i=A_{i-1}+\pi_i,\nonumber \\
&&\pi_i=\left\{\begin{array}{l}
+1 \mbox{;  nucleon from the left involved,}\\
-1 \mbox{;  nucleon from the right involved,}
\end{array} \right.\cr
&&\qquad  i=1,2,\cdots,\quad A_0=A, \label{eqA-1}
\end{eqnarray}
where $i$ counts the number of nucleon transfer processes after $t_0$.
The difference between the kinetic energy at the initial time $t_0$ and that
after the $i$-th nucleon transfer process is then expressed as
$T_{\rm diss}(i)=P_0^2/2mA-P_i^2/2mA_i$.
We denote the number of nucleons which have been exchanged up to a given relative
distance $R$ as $N_\mathrm{ex}(R)$. Then one-half of the total dissipation energy
of the collective motion at $R$ is expressed as
\begin{eqnarray}
T_{\rm diss}(R)&=&\frac{P_0^2}{2mA}-\frac{P_{N_\mathrm{ex}(R)}^2}{2mA_{N_\mathrm{ex}(R)}},\nonumber\\
P_{N_\mathrm{ex}(R)}&=&P_0-\sum_{i=1}^{N_\mathrm{ex}(R)}p_i,\nonumber\\
A_{N_\mathrm{ex}(R)}&=&A_0+\sum_{i=1}^{N_\mathrm{ex}(R)} \pi_i.  \label{eqA-2}
\end{eqnarray}
Here, it should be mentioned that the sum of the dissipation energy in
Eq.~(\ref{eqA-2}) and the amount of increased energy in the intrinsic
kinetic energy of the right (left) part is easily proven to be zero.
Namely, the amount of energy lost from the collective motion through the
nucleon exchange is just the same amount as that is increased in the intrinsic system.

\acknowledgements
We thank Lu Guo, K. Washiyama, En-Guang Zhao, and Yi-Zhong Zhuo
for helpful discussions.
This work has been supported by
the National Key Basic Research Program of China (Grant No. 2013CB834400),
the National Natural Science Foundation of China (Grants
No. 11075215,
No. 11121403,
No. 11120101005,
No. 11275052,
No. 11275248,
and
No. 11375062),
and
the Knowledge Innovation Project of the Chinese Academy of Sciences (Grant No. KJCX2-EW-N01).
F. S. appreciates the support by the Chinese Academy of
Sciences (CAS) Visiting Professorship for Senior
International Scientists (Grant No. 2013T1J0046).
The computational results presented in this work have been obtained on
the High-performance Computing Cluster of SKLTP/ITP-CAS and
the ScGrid of the Supercomputing Center, Computer Network Information Center of
the Chinese Academy of Sciences.



%

\end{document}